\documentclass[apj,twocolumn]{openjournal}
\usepackage{amsmath}
\usepackage{booktabs}
\usepackage{multirow}
\usepackage{color}
\usepackage{soul}

\usepackage[dvipsnames]{xcolor} 

\usepackage[breaklinks,colorlinks,citecolor=blue,urlcolor=blue]{hyperref}

\defcitealias{Shen2018}{S18}

\renewcommand{\arraystretch}{1.2}  
\usepackage{listings}
\usepackage{color}
\definecolor{dkgreen}{rgb}{0,0.6,0}
\definecolor{gray}{rgb}{0.5,0.5,0.5}
\definecolor{mauve}{rgb}{0.58,0,0.82}
\definecolor{golden}{rgb}{0.86,0.65,0.01}
\begin{document}


\title{On the formation of a $33\,M_{\odot}$ black hole in a low-metallicity binary}

\author{\vspace{-1cm}Kareem El-Badry$^{1}$}

\affiliation{$^1$Department of Astronomy, California Institute of Technology, 1200 E. California Blvd., Pasadena, CA 91125, USA}
\email{Corresponding author: kelbadry@caltech.edu}

\begin{abstract}
A $33\,M_\odot$ black hole (BH) was recently discovered in an 11.6-year binary only 590 pc from the Sun. The system, Gaia BH3, contains a $0.8\,M_\odot$ low-metallicity giant ($\rm [M/H]=-2.2$) that is a member of the ED-2 stellar stream. This paper investigates whether the system could have formed via isolated binary evolution. I construct evolutionary models for metal-poor massive stars with initial masses ranging from $35-55\,M_{\odot}$, which reach maximum radii of $1150-1800\,R_{\odot}$ as red supergiants. I then explore what combinations of initial orbit, mass loss, and natal kick can produce the period and eccentricity of Gaia BH3. Initial orbits wide enough to accommodate the BH progenitor as a red supergiant can match the observed period and eccentricity, but only if the BH formed with a significant natal kick ($v_{\rm kick}\gtrsim 20\, {\rm km\,s^{-1}}$). These models are disfavored because such a kick would have ejected the binary from the ED-2 progenitor cluster. I conclude that Gaia BH3 likely formed through dynamical interactions, unless the BH progenitor did not expand to become a red supergiant.  Only about 1 in 10,000  stars in the solar neighborhood have metallicities as low as Gaia BH3. This suggests that BH companions are dramatically over-represented at low-metallicity, though caveats related to small number statistics apply. The fact that the luminous star in Gaia BH3 has been a giant -- greatly boosting its detectability -- only for $\sim$1\% of the time since the system's formation implies that additional massive BHs remain to be discovered with only moderately fainter companions. Both isolated and dynamically-formed BH binaries with orbits similar to Gaia BH3 are likely to be discovered in Gaia DR4. 


\end{abstract}

\maketitle

\section{Introduction}
\label{sec:intro}
\citet{Panuzzo2024} recently reported discovery of an astrometric binary containing a $0.8\,M_{\odot}$ star on the lower giant branch and a $33\,M_\odot$ dark companion that is presumably a black hole (BH). The system, which they named Gaia BH3, was discovered with pre-release DR4 astrometry, and its orbital motion was confirmed with (nearly) independent radial velocity measurements from the {\it Gaia} RVS spectrometer \citep{Cropper2018}. 

The discovery is exciting for several reasons. First, Gaia BH3 contains the most massive robustly measured stellar-mass BH in the local Universe. Second, the star has the lowest metallicity of any known star orbiting a BH, $[\rm M/H]=-2.2$. Third, the system provides the first empirical evidence that low-metallicity massive stars leave behind massive BHs, making them prime candidates for being progenitors of gravitational wave sources. And fourth, the system is bright and nearby ($G=11.2$, $d=590$\,pc), suggesting that many similar systems that are fainter and/or more distant remain to be discovered. 

Gaia BH3 is the widest BH binary discovered to date, with an orbital period of $\sim 4250$ days and a semimajor axis of $\sim 16.5$\,au. The eccentricity is relatively high, $e\approx 0.73$, such that the star and BH pass within 4.5 au of each other at periastron. How the system formed is uncertain. \citet{Panuzzo2024} note that a typical red supergiant would not fit within the current orbit at periastron. They thus consider formation of the system from an isolated binary unlikely, and invoke dynamical exchange as a possible alternative. \citet{Balbinot2024} report that Gaia BH3 is a member of the ED-2 stellar stream \citep{Dodd2023}, lending credence to the dynamical formation hypothesis. Because ED-2 stars do not display light element abundance variations characteristic of globular clusters, \citet{Balbinot2024} propose that ED-2 formed from a low-mass cluster ($M_{\rm cluster} \lesssim 4\times 10^4\,M_{\odot}$).

Here, I consider the possibility of formation from an isolated binary in more detail. While the association of Gaia BH3 with ED-2 appears robust, the progenitor cluster's initial density and mass are uncertain, leaving open the possibility that Gaia BH3 is a primordial binary as opposed to one formed by dynamical exchange. I explore (a) the expected mass and maximum radius of the BH progenitor, and (b) the range of possible orbits that could have produced the current orbit in the presence of possible impulsive mass loss and kicks to the BH during its formation. I ultimately conclude that, while orbits similar to Gaia BH3's {\it can}  naturally be produced from isolated binaries if BHs are born with modest natal kicks, this channel is disfavored for Gaia BH3 in particular because such natal kicks would have ejected the binary from the ED-2 progenitor cluster.

The remainder of this paper is organized as follows. In Section~\ref{sec:mesa}, I describe evolutionary models for low-metallicity massive stars representative of the BH progenitor. Section~\ref{sec:kicks} uses Monte Carlo simulations of BH formation to assess the likely initial separation if Gaia BH3 formed from an isolated binary.  Section~\ref{sec:feh} argues that BH companions appear to be strongly overrepresented at low metallicity, and that additional nearby BHs orbited by low-metallicity stars likely remain to be discovered. Future prospects and broader implications are briefly discussed in Section~\ref{sec:disc}.

\section{The BH progenitor}
\label{sec:mesa}

\begin{figure*}
    \centering
    \includegraphics[width=\textwidth]{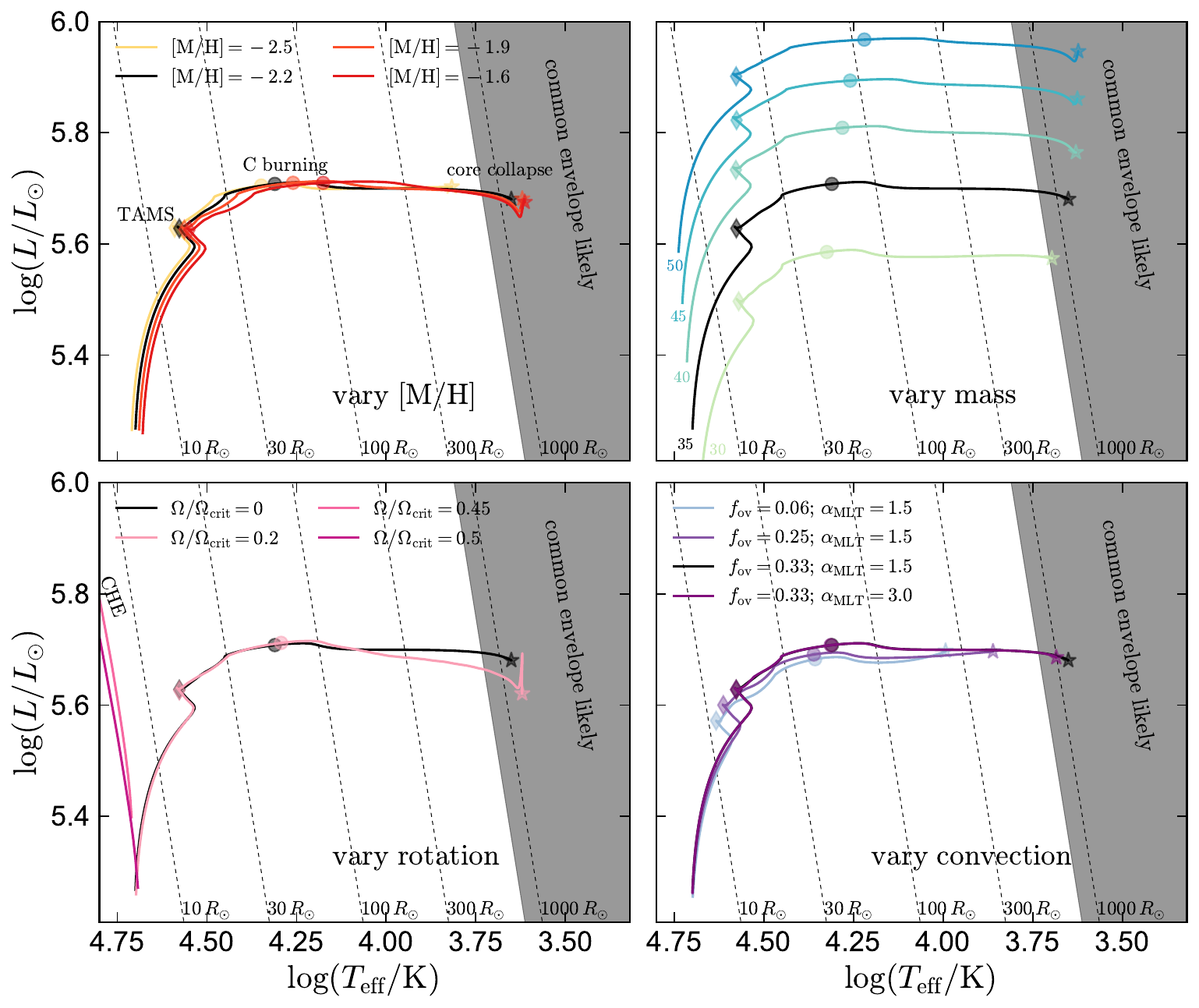}
    \caption{MESA evolutionary models for low-metallicity massive stars. The model shown in black is the same in all panels: a $35\,M_{\odot}$ star with metallicity $\rm [M/H] = -2.2$ and no rotation. Colored lines show models varying metallicity (upper left), initial mass (upper right), initial rotation rate (lower left), and overshooting or mixing length (lower right). Diamonds, circles, and stars mark core hydrogen exhaustion, the onset of carbon burning, and core collapse. Dashed diagonal lines show constant radius. Shaded gray region marks $R>800\,R_{\odot}$, beyond which the star would have been unlikely to fit inside the pre-SN orbit of Gaia BH3 (Section~\ref{sec:kicks}). Most of the models exceed this limit at their maximum radius and thus would have overflowed their Roche lobes. The only exceptions are rapidly-rotating models that undergo chemically homogeneous evolution (CHE; lower left) and models with reduced overshooting (lower right). }
    \label{fig:mesa}
\end{figure*}

To estimate the maximum radius and expected mass loss history of the BH progenitor, I calculated a small suite of evolutionary models using MESA \citep{Paxton2011, Paxton2013, Paxton2015, Paxton2018, Paxton2019, Jermyn2023}. The calculations closely follow the setup described by \citet{Klencki2020}, with MESA updated to version \texttt{r23.05.1}. The most important ingredients are summarized below, and I refer to \citet{Klencki2020, Klencki2021} for more details. 

I evolve single stars with initial masses ranging from 30 to $55\,M_{\odot}$ and metallicities ranging from $Z= 0.00002$ ($\rm [M/H] = -2.8$) to $Z = 0.0014$ ($\rm [M/H] = -1.0$). Convection is modeled following \citet{Henyey1965} (\texttt{mlt\_option = `Henyey'}) with a default mixing length parameter $\alpha=1.5$. The Ledoux criterion is used and  semiconvection is modeled following \citet{Langer1983}, with efficiency parameter $\alpha_{\rm SC} = 100$ motivated by \citet{Schootemeijer2019}. MLT++ is not used.  Step overshooting is assumed above the H and He burning cores, with \texttt{overshoot\_f = 0.33} and \texttt{overshoot\_f0 = 0.05} by default. I experiment with varying the mixing length and overshooting parameter, as described below.

Stellar winds are modeled following \citet{Brott2011}: for hot, hydrogen-rich stars ($T_{\rm eff} > 25$\,kK; $X_S>0.7$), the models take the wind mass loss rates from \citet[][here $X_S$ is the surface hydrogen mass fraction]{Vink2000, Vink2001}. For hydrogen-poor stars ($X_S < 0.4$), they use  the Wolf-Rayet wind model from \citet{Hamann1995} and reduce the mass loss rate by a factor of 10 to account for wind clumping \citep{Yoon2006}. For stars with intermediate surface hydrogen abundances ($0.4 < X_S < 0.7$), the mass loss rate is interpolated between the two prescriptions above. For stars cooler than $T_{\rm eff} = 25$\,kK, the mass loss rate is set to the larger of the rates predicted by \citet{Nieuwenhuijzen1990} and \citet{Vink2001}. These mass loss rates all include a $\dot M \propto Z^{0.85}$ scaling with metallicity. As a result, the predicted wind mass-loss rates are all rather low at the metallicities of interest for Gaia BH3. For consistency with \citet{Klencki2020}, the models assume $Z_\odot = 0.017$ when calculating mass loss rates, but I label the metallicities by $[{\rm M/H}] = \log(Z/0.014)$ following \citet{Asplund2009}. 

For rotating models, I include the effects of Eddington-Sweet circulation (\texttt{D\_ES\_factor = 1.0}), secular shear instabilities (\texttt{D\_SSI\_factor = 1.0}), and the Goldreich-Schubert-Fricke instability (\texttt{D\_GSF\_factor = 1.0}), with an efficiency factor \texttt{am\_D\_mix\_factor = 1/30} following \citet{Heger2000}. I initialize models using the MESA \texttt{create\_pre\_main\_sequence\_model} routine and run until the end of carbon burning, by which time the model is within days of core collapse and there is no time for further radius evolution. I limit the timestep with \texttt{varcontrol\_target = 0.0001} and \texttt{delta\_HR\_limit = 0.0005} and the mesh resolution with \texttt{max\_dq = 0.001}. Sensitivity to these choices was explored by \citet{Klencki2021}.

To verify that the problem setup is consistent with the one used by \citet{Klencki2020}, I began by reproducing their $35\,M_{\odot}$ model at metallicity $Z = 0.00017$ and comparing to their published tracks. The model reaches a maximum radius of $1331\,R_{\odot}$ shortly before core collapse. Their corresponding track has a maximum radius $1340\,R_{\odot}$ at the same point, and its evolution in the HR diagram is nearly indistinguishable from that of the model calculated here.

Figure~\ref{fig:mesa} shows the evolution of several models in the HR diagram. Core hydrogen exhaustion, the onset of carbon burning, and the end of carbon burning are marked in each panel with a diamond, circle, and star symbol, respectively. As a fiducial model -- shown in black in each panel -- I take the nonrotating model with initial mass of $35\,M_\odot$ and metallicity $\rm [M/H]=-2.2$. The mass of this model at the end of carbon burning is $34.7\,M_{\odot}$, reflecting the fact that winds are predicted to be very weak at these metallicities. This model could be expected to produce a $\sim 33\,M_{\odot}$ BH if the SN shock fails and it undergoes core collapse with little mass loss, since loss of up to a few $M_\odot$ is still expected to occur as a result of neutrino mass loss and the subsequent shock \citep{Fernandez2018}. The fiducial model reaches a maximum radius of $1150\,R_{\odot}$ shortly before core collapse. 

The four panels in Figure~\ref{fig:mesa} show the effects of varying metallicity, mass, rotation, and convective mixing length and overshooting. The maximum radius reached by our models varies relatively weakly with metallicity over $\rm -2.2 < [M/H] < -1.6$. The model with $\rm [M/H] = -2.5$ (slightly lower than Gaia BH3) is significantly smaller than other models, reaching a maximum radius of $\sim 550\,R_{\odot}$. The maximum radius increases with initial mass, ranging from  $\approx 850\,R_{\odot}$ for the $30\,M_{\odot}$ model to $\approx 1800\,R_{\odot}$ for the $50\,M_{\odot}$ model at $\rm [M/H]=-2.2$. The maximum radius is sensitive to overshooting, as explored by \citet{Schootemeijer2019}: smaller overshooting parameters lead to more compact supergiants. For the choice of parameters explored here, sensitivity to the mixing length is relatively weak: increasing $\alpha_{\rm MLT}$ from 1.5 to 3.0 reduces the maximum radius by $\sim 10\%$.  

The shaded regions in Figure~\ref{fig:mesa} show $R>800\,R_{\odot}$. In Section~\ref{sec:kicks}, I argue that this is the largest progenitor that could plausibly have fit inside the orbit of Gaia BH3 before the BH's formation. Most of the models exceed this radius shortly before core collapse. A notable exception is provided by the two rapidly-rotating models shown in the lower left panel of Figure~\ref{fig:mesa}, which go through chemically homogeneous evolution (CHE). At high rotation rates, large-scale meridional circulations can efficiently mix helium into stellar envelopes, preventing the buildup of a chemical gradient and core/envelope dichotomy that leads to expansion in nonrotating models \citep{Maeder1987, Yoon2005, Woosley2006}. Under CHE, stars evolve from the main sequence toward the helium main sequence, where they have effective temperatures above $10^5\,\rm K$ and radii smaller than $1\,R_\odot$. These models eventually ignite carbon burning and terminate their evolution without ever reaching $R>10\,R_{\odot}$. At the low metallicities considered here, CHE is predicted to occur with initial rotation rates of $\Omega/\Omega_{\rm crit} \geq 0.45$.

\begin{figure}
    \centering
    \includegraphics[width=\columnwidth]{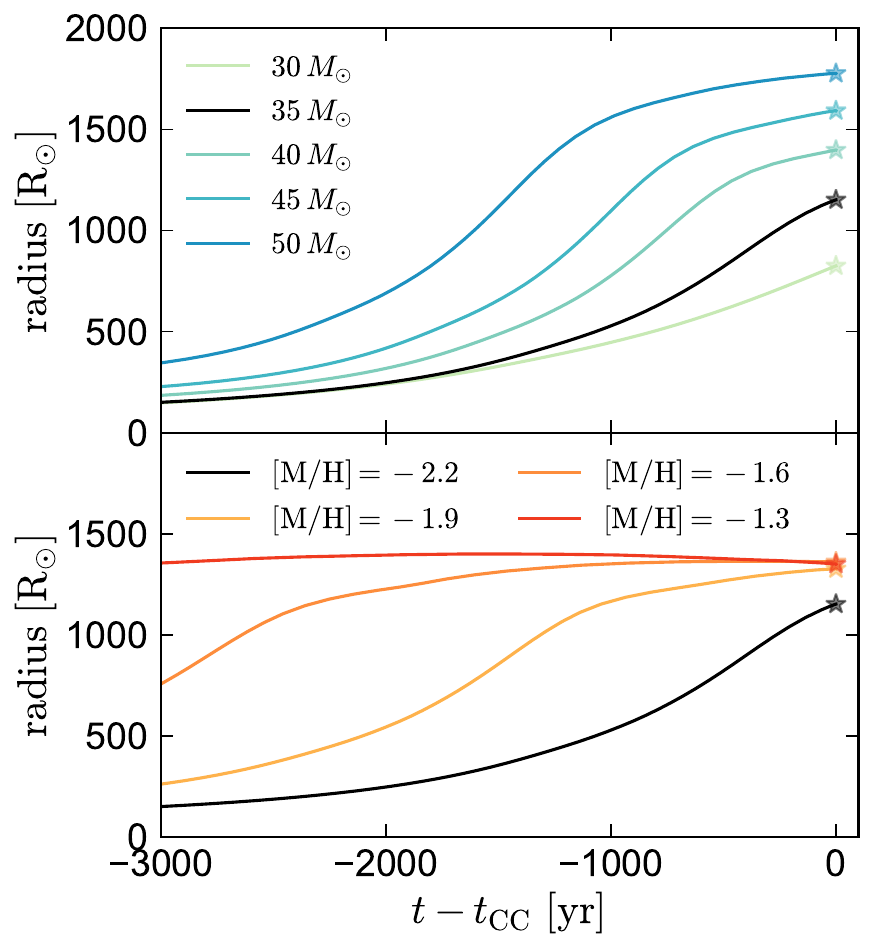}
    \caption{Radius evolution of MESA models in the final years before core collapse. Mass is varied in the top panel, and metallicity in the bottom panel. At low metallicity, the models expand significantly in the final $\sim 1000$ yr of their evolution, only appearing as red supergiants shortly before core collapse. Expansion occurs earlier at higher metallicity. This means that wide low-metallicity binaries are less likely to be tidally circularized before the formation of a BH. }
    \label{fig:rad_evol}
\end{figure}

One of the main differences between the models shown here and higher-metallicity massive star models is that low-metallicity models remain compact during core helium burning: although they are red supergiants just before core collapse, they spend the large majority of their post-main sequence evolution as blue or yellow supergiants with $R\lesssim 100\,R_\odot$, only expanding to red supergiant dimensions during their final $\sim 1000$ years \citep[see][for further discussion]{Klencki2020}. This is illustrated in Figure~\ref{fig:rad_evol}, which shows the late-stage radius evolution for models with a range of masses (top panel) and metallicities (bottom panel). The fiducial model increases its radius by a factor of 2 in its final 800 years. This behavior is not found in higher metallicity models, even at ``low'' metallicites comparable to those found in the SMC and LMC. The upshot is that if Gaia BH3 formed as a primordial binary, there may not have been sufficient time for tides to circularize the orbit when the BH progenitor was a red supergiant. 
 
\section{Constraints on the progenitor orbit}
\label{sec:kicks}
I use Monte Carlo simulations to study the combinations of pre-supernova\footnote{Here ``supernova'' loosely refers to the death of the massive star and formation of the BH. Whether there was a successful explosion or a long-lived luminous transient is uncertain.} (SN) orbits and natal kicks that could have produced an orbit similar to that of Gaia BH3. This approach follows \citet{Tauris2017} and \citet{El-Badry2024_ns1}; see \citet{Tauris2023} for additional details about SN kicks. In brief, I simulate a large number ($N=10^7$) of orbits and kicks, assume that the SN occurs at a random phase with a kick oriented in a random direction, and predict post-SN parameters for all orbits. Finally, I analyze the initial parameters of the simulations for which the final period and eccentricity are close to the observed values for Gaia BH3: $P_{\rm orb, final} = 4000-4500$\,d, and $e_{\rm final} = 0.68-0.78$. Narrowing these ranges reduces the number of surviving Monte Carlo samples but has no significant effect on their distribution.

\begin{figure*}[!ht]
    \centering
    \includegraphics[width=\textwidth]{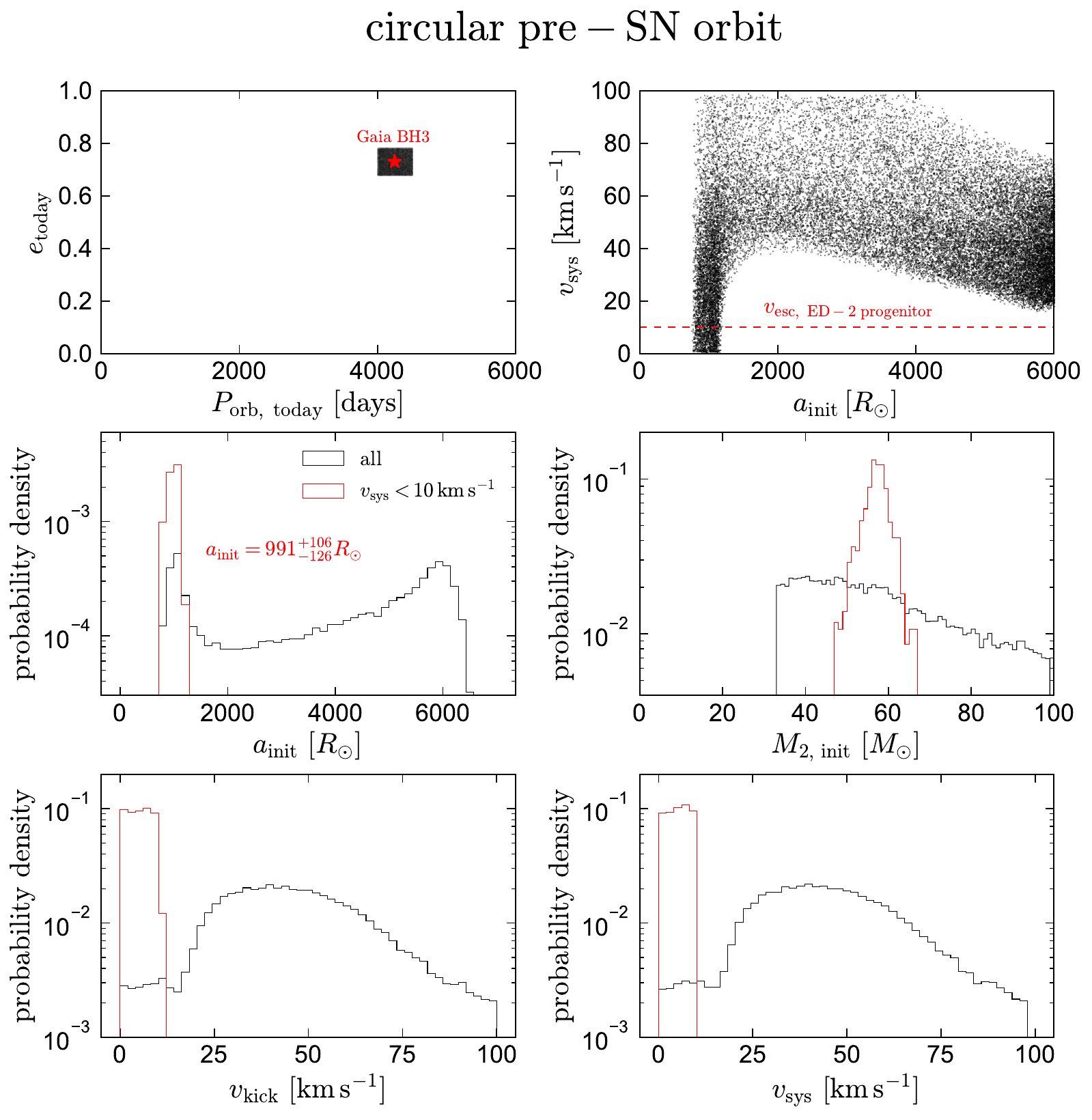}
    \caption{Constraints on kicks and the pre-SN orbit of Gaia BH3-like binaries. I assume that the binary had a circular orbit before the SN and that the BH received a kick with velocity $v_{\rm kick}$ when it formed. I simulate $10^7$ pre-SN orbits with a uniform period distribution $P_{\rm orb} \sim \mathcal{U}(0,2P_{\rm orb,BH3})$, a pre-SN mass distribution $M_{2,\rm \,init}/M_\odot \sim \mathcal{U}(33,100)$, and a kick velocity distribution $v_{\rm kick}/({\rm km\,s^{-1}}) \sim \mathcal{U}(0,100)$. Black histograms show the properies of all the systems whose post-SN orbits have periods and eccentricities similar to Gaia BH3 (upper left). Red histograms show the subset of these that result in a post-SN systemic velocity $v_{\rm sys} <10\,\rm km\,s^{-1}$, as required to remain bound to the ED-2 stream progenitor. Orbits with a broad range of pre-SN separations can produce the observed period and eccentricity. However, most of these require significant kicks and result in $v_{\rm sys} >10\,\rm km\,s^{-1}$. Simulations with $v_{\rm sys} < 10\,\rm km\,s^{-1}$ can only produce the observed period and eccentricity for $a_{\rm init} \lesssim 1200\,R_{\odot}$. The BH progenitor would overflow its Roche lobe in such orbits for $R\gtrsim 800\,R_{\odot}$. }
    \label{fig:kicks}
\end{figure*}

\subsection{Circular orbits}
\label{sec:circular}
I first consider pre-SN orbits that have zero eccentricity, as would be expected if tides circularized the orbit when the BH progenitor expanded. In this case, the system's current eccentricity must be a result of natal kicks and/or instantaneous mass loss during the SN. I use the formalism from \citet{Brandt1995} to predict post-SN parameters. 

I begin with a uniform distribution of pre-SN orbital periods between 0 and 8500\,d (twice the currently observed orbital period of Gaia BH3). I simulate kicks assuming orientations distributed uniformly on the sky and velocities distributed uniformly between 0 and 100\,$\rm km\,s^{-1}$. I take a uniform distribution of $\Delta m$, the mass loss of the BH progenitor during the SN, between 0 and 67\,$M_{\odot}$, implying a progenitor mass of 33-100\,$M_{\odot}$. 
This is not a population synthesis simulation -- pre-SN orbits are unlikely to be uniformly distributed in all parameters -- but should be viewed as an exploration of what combinations of orbits, kicks, and mass loss could have produced the observed orbit.

The results of this experiment are shown in Figure~\ref{fig:kicks}. Black histograms show all simulations that produce post-SN periods and eccentricities similar to Gaia BH3 today, while red histograms show the subset of those that predict a post-SN systemic velocity $v_{\rm sys} < 10\,\rm km\,s^{-1}$. This is a conservative estimate of the largest post-SN systemic velocity for which Gaia BH3 could plausibly have remained bound to the low-mass cluster that formed the ED-2 stream. 

When there is no restriction on $v_{\rm sys}$ (black histograms), initial separations ranging from $\sim 800$ to $\sim 6000\,R_\odot$ can produce the observed orbit. Initial separation closer than $\sim 800\,R_{\odot}$ cannot, because widening these to the separation observed today via a kick or mass loss would result in a higher final eccentricity than is observed. Wider initial separations {\it could} match the orbit for suitably chosen kicks, but these are excluded by the prior of $P_{\rm orb,\,init} < 8500$\,d.  

The upper right panel of Figure~\ref{fig:kicks} shows the pre-SN separations and post-SN systemic velocities of all simulations that match the observed period and eccentricity. The combination of low $v_{\rm sys}$ ($\lesssim 20\,\rm km\,s^{-1}$) and large initial separation ($\gtrsim 1200\,R_{\odot}$) is ruled out because such models produce too low eccentricities.  The red histograms show that among simulations with $v_{\rm sys} < 10\,\rm km\,s^{-1}$, only relatively tight pre-SN orbits with $a_{\rm init} < 1200\,R_{\odot}$ can match the observed period and eccentricity. These calculations all have weak kicks  ($v_{\rm kick} \lesssim 10\,\rm km\,s^{-1}$) and substantial mass loss, such that the final eccentricity is mainly a result of the \citet{Blaauw1961} kick. Initial masses $M_{2,\rm init} \gtrsim 70\,M_{\odot}$ are ruled out because they become unbound due to mass loss. 

The largest star that could fit inside a pre-SN orbit of semimajor axis $a_{\rm init}$ has radius $R_{\rm max} = a_{\rm init} f_q$, where $f_q(M_1/M_2)$ is given by \citet{Eggleton1983} and varies from 0.68 to 0.71 over $M_1 = 35-70\,M_{\odot}$, assuming $M_2 = 0.8\,M_{\odot}$. This means that for the pre-SN orbits that can produce the observed period and eccentricity of Gaia BH3 without escaping the ED-2 progenitor, the $\pm 1\sigma$ range of maximum radii is $R_{\rm max} \approx  680_{-87}^{+73}\,R_{\odot}$. As shown in Section~\ref{sec:mesa}, most plausible models for the progenitor of the BH reach maximum radii larger than this value, so it seems unlikely that the present orbit of Gaia BH3 could have resulted from isolated binary evolution if the pre-SN orbit was circular. 

\subsection{Eccentric orbits}
\label{sec:eccentric}

\begin{figure*}
    \centering
    \includegraphics[width=\textwidth]{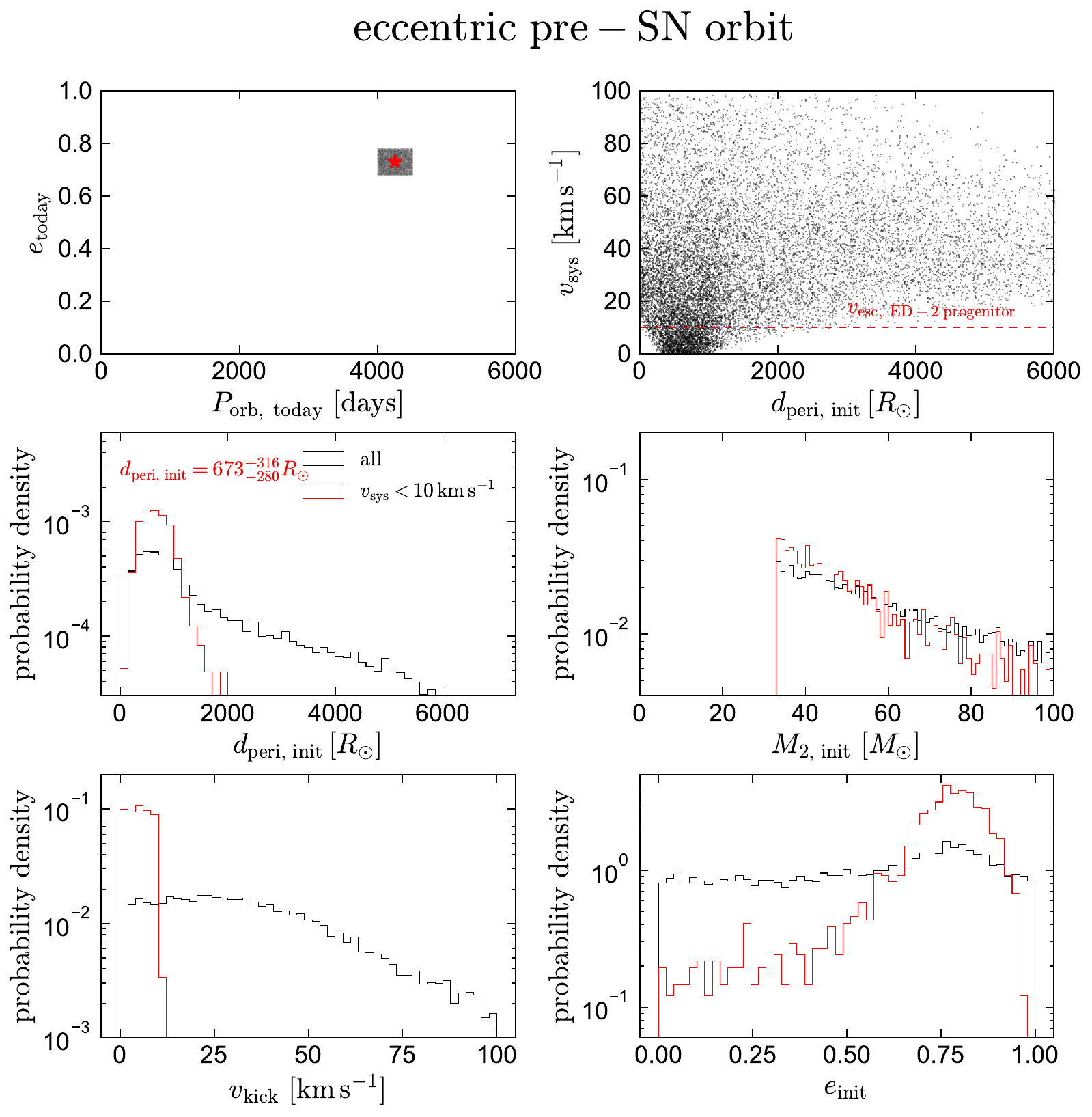}
    \caption{Similar to Figure~\ref{fig:kicks}, but for eccentric pre-SN orbits. I simulate $10^7$ orbits with initial eccentricity $e_{\rm init}\sim \mathcal{U}(0,1)$ and the SN occurring at a random phase. Other parameters are distributed as in Figure~\ref{fig:kicks}. $d_{\rm peri,\,init}$ is the separation between the BH progenitor and companion star at periastron in the pre-SN orbit. Most orbits with $v_{\rm sys} < 10\,\rm km\,s^{-1}$ (red) have $d_{\rm peri,\,init} \lesssim 1100\,R_{\odot}$ and thus would overflow their Roche lobes if the BH progenitor reached $R\gtrsim 800\,R_{\odot}$.  }
    \label{fig:kicks_ecc}
\end{figure*}

I next consider pre-SN orbits that are eccentric. These may be expected if the BH progenitor only expands late in its evolution, leaving insufficient time for tidal circularization (e.g. Figure~\ref{fig:rad_evol}). I again simulate $N=10^7$ orbits, now with a uniform distribution of eccentricities between 0 and 1. I assume that the SN occurs at a random phase, corresponding to a uniform distribution of the pre-SN mean anomaly, and use the formalism from \citet{Hurley2002} to predict the post-SN orbital eccentricity, period, and center-of-mass velocity. 

The results are shown in Figure~\ref{fig:kicks_ecc}. In this case, I show the distributions of pre-SN periastron distance, $d_{\rm peri,\,init}$, rather than semimajor axis, since this is the key parameter for determining whether the BH progenitor would have overflowed its Roche lobe. As in the circular case, a broad range of initial orbits are possible for arbitrary kicks, but a relatively narrow range of $d_{\rm peri,\,init}$ can produce the observed period and eccentricity subject to the restriction of $v_{\rm sys} < 10\,\rm km\,s^{-1}$. Most of these orbits have weak kicks and little mass loss, and thus are similar to the orbit of Gaia BH3 today, which has $d_{\rm peri} \approx 800\,R_{\odot}$. Orbits with significantly larger $d_{\rm peri,\,init}$ and weak kicks end up with longer periods and/or lower eccentricities than the values observed for Gaia BH3 today.

Among orbits that reproduce the observed eccentricity and period of Gaia BH3 with $v_{\rm sys} < 10\,\rm km\,s^{-1}$, 90\% have $d_{\rm peri,\,init}< 1120\,R_{\odot}$, such that they would overflow their Roche lobes at periastron for radii $R>775\,R_{\odot}$. 99\% would overflow their Roche lobes at periastron for $R>1125\,R_{\odot}$. Given that the MESA models shown in Figure~\ref{fig:mesa} predict maximum radii of $\gtrsim 1000\,R_{\odot}$, I conclude that the BH progenitor is unlikely to have avoided Roche lobe overflow in any orbit that could have produced the current orbit of Gaia BH3, unless it formed through chemically homogeneous evolution or had a maximum radius significantly smaller than predicted by the evolutionary models shown in Figure~\ref{fig:mesa}.

\section{Population inference from $N = 1$}
\label{sec:feh}

\subsection{Wide BH companions are overrepresented at low metallicity}
\begin{figure}
    \centering
    \includegraphics[width=\columnwidth]{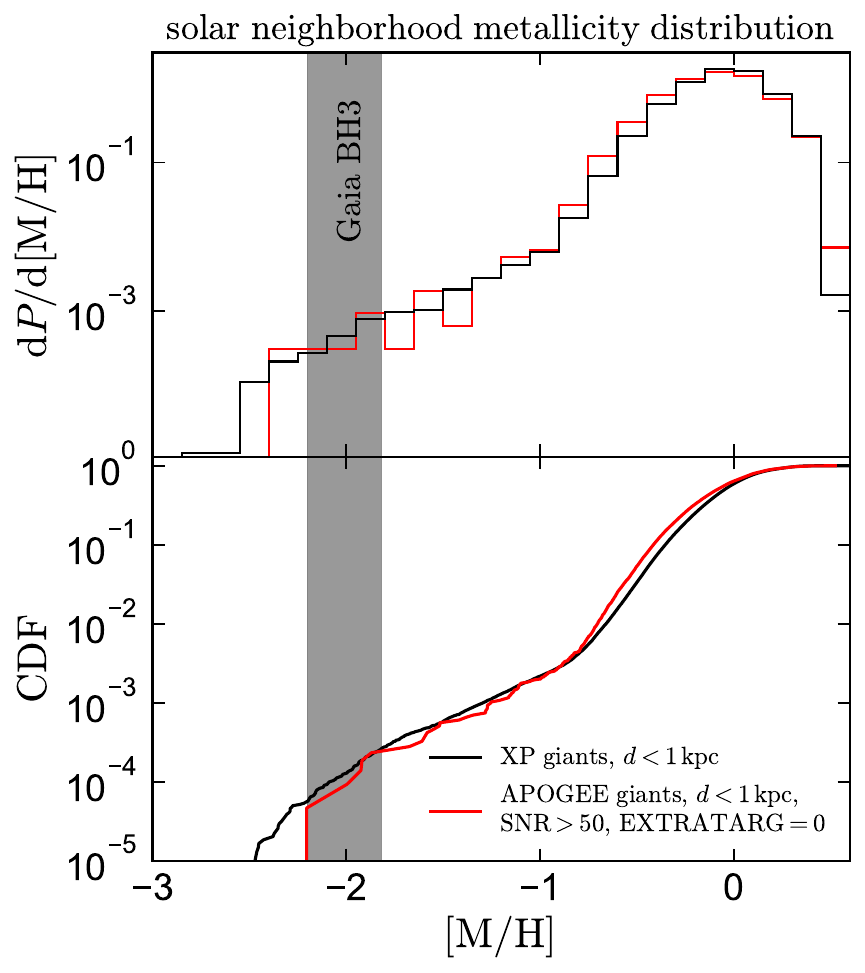}
    \caption{Metallicity distribution of red giants within 1\,kpc of the Sun, calculated from the {\it Gaia} XP metallicity catalog of \citet[][black]{Andrae2023} and APOGEE DR17  \citep[][red]{Abdurro'uf2022}. The metallicity of Gaia BH3 is marked with a shaded region: $\rm [M/H]=-1.82$ according to \citet{Andrae2023}, and $\rm [M/H]=-2.2$  according to \citet{Panuzzo2024}. The fraction of giants with metallicities as low as Gaia BH3 is only $\sim 0.0001$. }
    \label{fig:mdf}
\end{figure}

Gaia BH3 has a much lower metallicity than typical stars in the solar neighborhood. It is expected that low-metallicity massive stars may form higher-mass BHs due to their weaker winds \citep{Woosley2002}, and Gaia BH3 was published in advance of DR4 primarily because of the BH's high mass. However, I argue here that -- even ignoring the BH's mass -- the fact that a BH of any mass was found with such a low-metallicity companion suggests that wide BH companions are overrepresented at low metallicity. 

Figure~\ref{fig:mdf} shows differential and cumulative metallicity distributions of giants within 1 kpc of the Sun. In black, I show giants with high-quality metallicities from {\it Gaia} XP spectra, which I take from Table 2 of \citet{Andrae2023}. In red, I show giants with high-SNR spectra from DR 17 of the APOGEE survey \citep{Majewski2017, Abdurro'uf2022}. Here I exclude stars targeted non-randomly, since several of the APOGEE targeting cartons intentionally selected low-metallicity stars. The shaded gray region shows the plausible metallicity uncertainty of Gaia BH3, which goes from $\rm [M/H]=-2.2$ as estimated by \citet{Panuzzo2024} to $\rm [M/H]=-1.82$ as estimated for the source by \citet{Andrae2023}. I assume $\rm [M/H]=-2.0$ in the discussion below.

Both the XP and APOGEE samples suggest that only 1 in $\sim 10^4$ giants within 1 kpc have $\rm [M/H]<-2.0$. This very small fraction strongly implies that wide BH companions are overrepresented at low metallicity. If they were not, we could expect to find $\sim10^4$ BH companions -- just to red giants -- within 1 kpc. The true number of BH + giant binaries within 1 kpc is unknown, but 0 have been discovered so far: the nearest such system is Gaia BH2 at $d=1.16$\,kpc \citep{El-Badry2023_bh2}. It seems quite unlikely that $10^4$ remain to be discovered (although published {\it Gaia} data are not yet sensitive to orbital periods as long as that of Gaia BH3), and so a more plausible explanation is that wide BH companions are more common at low metallicity. Early results from searches for neutron star companions in the {\it Gaia} data similarly imply that they are overrepresented at low metallicity \citep{El-Badry2024_nsall}. More work is needed to understand why.

\subsection{More low-metallicity BHs are expected in DR4}
\label{sec:moar}
There is also reason to believe that more BHs orbited by metal-poor stars exist in the solar neighborhood. Such stars are uniformly $\gtrsim 12$ Gyr old. This means that giants sample a narrow mass range of roughly $(0.78-0.80)\,M_{\odot}$. For a Kroupa IMF, there are $\sim$165 stars with $M = (0.1-0.78)\,M_{\odot}$ for every star with $M=(0.78-0.80)\,M_{\odot}$, suggesting that even within 1 kpc, other massive BHs are likely orbited by lower-mass metal-poor stars that are still on the main sequence. The mass distribution of companions to BHs of course may be different from the IMF, but it seems unlikely that $0.8\,M_{\odot}$ companions would be significantly more common than companions with mass of $0.7\,M_{\odot}$ or $0.6\,M_{\odot}$. At the distance and extinction of Gaia BH3, such companions would have $G<14$ for $M>0.75\,M_{\odot}$, or $G < 16$ for $M>0.61\,M_{\odot}$. These magnitudes are too faint for {\it Gaia} to have measured multi-epoch RVs from RVS spectra, but bright enough that the astrometric orbit would be resolved with $\rm SNR \gtrsim 100$. With careful processing of the astrometric data, a source like Gaia BH3 could likely be detected even at $G=18-19$ -- where individual astrometric measurements have uncertainties of 1-3\,mas in the along-scan direction, a factor of 10-30 smaller than Gaia BH3's photocenter ellipse \citep{Lindegren2018, Lindegren2021, Holl2023} -- although spurious solutions will be more abundant at low SNR. Gaia BH3-like systems can thus quite plausibly be detected with significantly lower-mass secondaries, and to larger distances, than Gaia BH3 itself.

Unless Gaia BH3 is a major statistical fluke, these considerations suggest that (a) more massive BHs remain to be discovered around nearby metal-poor stars, and (b) the occurrence rate of BH companions is considerably higher at low metallicity than at solar values.

\section{Discussion}
\label{sec:disc}

\subsection{Formation from an isolated binary}
I have shown that wide BH + stellar binaries with orbits similar to Gaia BH3 can in principle be produced through isolated binary evolution, without the BH progenitor ever interacting with its companion. However, the orbit of Gaia BH3 in particular appears unlikely to have formed from this channel if we accept that the binary was born in the progenitor of the ED-2 stream and remained bound to it when the BH formed: matching the system's observed period and eccentricity while accommodating the BH progenitor as a red supergiant requires a natal kick of velocity $\gtrsim 20\,{\rm km\,s^{-1}}$, and such a kick would have ejected the binary from a low-mass cluster. Given the old age of the systems and the short lifetime of the BH progenitor, it seems unlikely that the binary was ejected from the stream with a significant kick and still remains closely associated with the stream in phase space 

The main uncertainty in these calculations is in the maximum radii predicted by the evolutionary models. For example, Figure~\ref{fig:mesa} shows that rapidly-rotating BH progenitors could have undergone CHE and never overflowed their Roche lobes in any plausible pre-SN orbit of Gaia BH3. The radii of red supergiants are also sensitive to a variety of uncertain inputs, such as the convective mixing length and the treatment of overshooting and semiconvection \citep{Chun2018, Schootemeijer2019, Goldberg2020, Klencki2021, Romagnolo2023}. Indeed, \citet{Iorio2024} perform similar calculations to those presented here but employ evolutionary tracks from the MIST project \citep{Choi2016}, which predict maximum radii of only $\sim 200\,R_{\odot}$.  Based on these models, they find that Gaia BH3's properties are consistent with formation from an isolated binary. The modeling choices adopted here produce red supergiant radii in reasonably good agreement with observations of massive stars in the LMC and SMC \citep[e.g.][]{Schootemeijer2019}, which are the lowest-metallicity galaxies currently observable with a large population of evolved massive stars. The maximum radii they predict at $\rm [M/H]=-2.2$ are not significantly different at fixed mass from those predicated at $\rm [M/H]=-0.7$, the metallicity of the SMC. No one has ever observed a massive star with a metallicity as low as Gaia BH3, so the evolutionary models are untested in the lowest-metallicity regime. 

\subsection{Dynamical formation}
With formation from a primordial binary unlikely, the alternative is dynamical assembly in the cluster that formed ED-2. Cluster formation models have also previously been found promising for forming the other two BH binaries discovered so far from {\it Gaia} data \citep{Rastello2023, DiCarlo2023, Tanikawa2024}, and for BH binaries found in globular clusters with spectroscopic surveys, some of which have similar parameters to the {\it Gaia} BHs \citep{Kremer2018, Gieles2021}.

Recent modeling by \citet{MarinPina2024} predicts a population of binaries with properties similar to Gaia BH3 that escape from Monte Carlo cluster models. They predict a separation distribution for dynamically-formed binaries that peaks at close separations but is relatively flat in log period over $P_{\rm orb} = 10^{0-4}$\,d. All the clusters in the library analyzed by \citet{MarinPina2024} have total masses significantly larger than \citet{Balbinot2024} estimate for the progenitor of the ED-2 stream, so it would be useful to model the formation of Gaia BH3-like binaries in lower-mass clusters. The most extensive effort in this direction thus far was undertaken by \citet{Tanikawa2024, Tanikawa2024b}, who also predict a period distribution that is relatively flat in log space for clusters with $M_{\rm cluster}=10^3\,M_{\odot}$. A full exploration of the low-mass cluster regime is still warranted.

\citet{MarinPina2024} predict that the formation efficiency of wide BH binaries per unit mass increases as $\sim M_{\rm cluster}^{-1.2}$, such that low-mass clusters will contribute the large majority of BH+star binaries and the predicted number of detectable binaries depends sensitively on the behavior of the adopted cluster mass function at the low-mass end. Integrating down to $M_{\rm cluster}=10^2\,M_{\odot}$ (which may be optimistic, since the lowest-mass clusters may not form any BHs or be dense enough for dynamical interactions to be important), they predict $\lesssim 1$ low-metallicity BH binary within 1 kpc. The empirical rate considerations in Section~\ref{sec:moar} suggests that the true number of such binaries may be larger than this, but the Poisson uncertainty associated with $N=1$ detection still allows for the possibility that Gaia BH3 was a lucky discovery.

\subsection{The possibility of pollution}
It may be possible to discriminate between dynamical and isolated binary evolution channels for future BH discoveries on the basis of the abundances of the companion stars. \citet{Panuzzo2024} and \citet{Balbinot2024} show that Gaia BH3's companion does not show any significant abundance anomalies, and they suggest that this favors a dynamical formation channel. In support of this possibility, some of the neutron star binary candidates identified by \citet{El-Badry2024_nsall} do show abundance anomalies, though their origin is still uncertain. 

For Gaia BH3, I find that pollution from SN ejecta would likely not be detectable, even if the system had formed from isolated binary evolution. For a plausible separation of 10\,au and companion star radius of $1\,R_{\odot}$ at the time of the SN, the companion would have subtended a fraction $f \sim 5\times 10^{-8}$ of the sky as viewed from the BH progenitor. Assuming $20\,M_{\odot}$ of isotropic ejecta, only $\sim 10^{-6}\,M_{\odot}$ of ejecta -- likely consisting mostly of oxygen \citep{Thielemann1996} -- would have been deposited on the star. Since the star is now a giant, its convective zone extends deep into the interior: from a MESA model, I find that the outer $\sim 0.35\,M_{\odot}$ is likely convective. If $\sim 7\times 10^{-7}\,M_{\odot}$ of oxygen had been accreted, the ejecta would have  been mixed with $\sim 3\times 10^{-5}\,M_{\odot}$ of oxygen already in the envelope, changing the star's surface abundances by at most a few percent, which is unlikely to be detectable. Chances of detecting pollution would be much higher if the companion star were still on the main sequence, where the outer convective zone is predicted to contain $<10^{-3}\,M_{\odot}$.

\subsection{Future prospects}

Gaia BH3 is just at the separation beyond which BH+star binaries can avoid a common envelope. Irrespective of whether Gaia BH3 in particular formed from an isolated binary, the models presented here imply that isolated binary evolution produces binaries with orbits similar to Gaia BH3, and future {\it Gaia} data releases are likely to find these systems. This makes the system different from Gaia BH1 and BH2, which most studies have found difficult to produce through isolated binary evolution under standard assumptions (\citealt{El-Badry2023_bh1, Rastello2023, Tanikawa2024}, though see \citealt{Kotko2024} for a different view).

Constraints on the separation  distribution of BH binaries from DR4 will provide further insights into their dominant formation channel. If these systems form mainly by isolated binary evolution, models predict the BH + low mass star period distribution to rise steeply at $P_{\rm orb} \gtrsim 10$ yr \citep[e.g.][]{Breivik2017, Chawla2022}, the period beyond which a significant fraction of binaries will have avoided a common envelope inspiral. For dynamical formation channels, the period distribution will depend on the mass of the clusters in which most systems are formed, but there is little reason to expect a steep increase at $P_{\rm orb} \gtrsim 10$ yr, and models published so far predict a period distribution that is flat or declines at longer periods \citep{Tanikawa2024, MarinPina2024}. Fortunately, {\it Gaia} will obtain an 11-year observational baseline \citep{GaiaCollaboration2016}, enabling robust orbital constraints at periods up to $\sim 20$ years, and perhaps even significantly longer \citep[e.g.][]{Andrews2023}. 

The metallicity distribution of BH binaries may also help constrain how they formed. The vast majority of clusters that have contributed stars to the solar neighborhood had metallicities near solar. If dynamical interactions form most wide BH binaries, most systems should likely have metallicities near solar, similar to Gaia BH1 and BH2 \citep{El-Badry2023_bh1, El-Badry2023_bh2}. For example, \citet{Tanikawa2024b} predict that the formation efficiency per unit mass of dynamically formed BH + star binaries increases by less than a factor of $\sim 5$ between $\rm [M/H] = 0$ and $\rm [M/H] = -2$. Curiously, the empirical data so far suggests a significant increase in the prevalence of both BH and neutron star companions at low metallicity \citep{El-Badry2024_ns1, El-Badry2024_nsall}.

We do not know how many other BHs in the pre-release {\it Gaia} data were {\it not} published. Because Gaia BH3 has too long of an orbital period to have been accessible in DR3 -- and the BH binary period distribution may rise steeply at long periods, where binaries can avoid interaction and a common envelope -- population inference is still uncertain.  But prospects for constraining this population in the coming years are bright.

\section*{acknowledgments}
I thank the anonymous referee for comments that significantly improved the paper. I am also grateful to Jim Fuller, Re'em Sari, Jared Goldberg, Thomas Tauris, Hans-Walter Rix, Vedant Chandra, and Rene Andrae for conversations that were influential to this work, and Jakub Klencki for making his MESA inlists publicly available. This research was supported by NSF grant AST-2307232.

This work has made use of data from the European Space Agency (ESA) mission
{\it Gaia} (\url{https://www.cosmos.esa.int/gaia}), processed by the {\it Gaia}
Data Processing and Analysis Consortium (DPAC,
\url{https://www.cosmos.esa.int/web/gaia/dpac/consortium}). Funding for the DPAC
has been provided by national institutions, in particular the institutions
participating in the {\it Gaia} Multilateral Agreement.


\bibliography{manuscript}

\begin{thebibliography}{}
\makeatletter
\relax
\def\mn@urlcharsother{\let\do\@makeother \do\$\do\&\do\#\do\^\do\_\do\%\do\~}
\def\mn@doi{\begingroup\mn@urlcharsother \@ifnextchar [ {\mn@doi@} {\mn@doi@[]}}
\def\mn@doi@[#1]#2{\def\@tempa{#1}\ifx\@tempa\@empty \href {http://dx.doi.org/#2} {doi:#2}\else \href {http://dx.doi.org/#2} {#1}\fi \endgroup}
\def\mn@eprint#1#2{\mn@eprint@#1:#2::\@nil}
\def\mn@eprint@arXiv#1{\href {http://arxiv.org/abs/#1} {{\tt arXiv:#1}}}
\def\mn@eprint@dblp#1{\href {http://dblp.uni-trier.de/rec/bibtex/#1.xml} {dblp:#1}}
\def\mn@eprint@#1:#2:#3:#4\@nil{\def\@tempa {#1}\def\@tempb {#2}\def\@tempc {#3}\ifx \@tempc \@empty \let \@tempc \@tempb \let \@tempb \@tempa \fi \ifx \@tempb \@empty \def\@tempb {arXiv}\fi \@ifundefined {mn@eprint@\@tempb}{\@tempb:\@tempc}{\expandafter \expandafter \csname mn@eprint@\@tempb\endcsname \expandafter{\@tempc}}}

\bibitem[\protect\citeauthoryear{{Abdurro'uf} et~al.,}{{Abdurro'uf} et~al.}{2022}]{Abdurro'uf2022}
{Abdurro'uf} et~al., 2022, \mn@doi [\apjs] {10.3847/1538-4365/ac4414}, \href {https://ui.adsabs.harvard.edu/abs/2022ApJS..259...35A} {259, 35}

\bibitem[\protect\citeauthoryear{{Andrae}, {Rix}  \& {Chandra}}{{Andrae} et~al.}{2023}]{Andrae2023}
{Andrae} R.,  {Rix} H.-W.,   {Chandra} V.,  2023, \mn@doi [\apjs] {10.3847/1538-4365/acd53e}, \href {https://ui.adsabs.harvard.edu/abs/2023ApJS..267....8A} {267, 8}

\bibitem[\protect\citeauthoryear{{Andrews}, {Breivik}, {Chawla}, {Rodriguez}  \& {Chatterjee}}{{Andrews} et~al.}{2023}]{Andrews2023}
{Andrews} J.~J.,  {Breivik} K.,  {Chawla} C.,  {Rodriguez} C.~L.,   {Chatterjee} S.,  2023, \mn@doi [\apj] {10.3847/1538-4357/acbb5f}, \href {https://ui.adsabs.harvard.edu/abs/2023ApJ...946..111A} {946, 111}

\bibitem[\protect\citeauthoryear{{Asplund}, {Grevesse}, {Sauval}  \& {Scott}}{{Asplund} et~al.}{2009}]{Asplund2009}
{Asplund} M.,  {Grevesse} N.,  {Sauval} A.~J.,   {Scott} P.,  2009, \mn@doi [\araa] {10.1146/annurev.astro.46.060407.145222}, \href {https://ui.adsabs.harvard.edu/abs/2009ARA&A..47..481A} {47, 481}

\bibitem[\protect\citeauthoryear{{Balbinot} et~al.,}{{Balbinot} et~al.}{2024}]{Balbinot2024}
{Balbinot} E.,  et~al., 2024, \mn@doi [arXiv e-prints] {10.48550/arXiv.2404.11604}, \href {https://ui.adsabs.harvard.edu/abs/2024arXiv240411604B} {p. arXiv:2404.11604}

\bibitem[\protect\citeauthoryear{{Blaauw}}{{Blaauw}}{1961}]{Blaauw1961}
{Blaauw} A.,  1961, \bain, \href {https://ui.adsabs.harvard.edu/abs/1961BAN....15..265B} {15, 265}

\bibitem[\protect\citeauthoryear{{Brandt} \& {Podsiadlowski}}{{Brandt} \& {Podsiadlowski}}{1995}]{Brandt1995}
{Brandt} N.,  {Podsiadlowski} P.,  1995, \mn@doi [\mnras] {10.1093/mnras/274.2.461}, \href {https://ui.adsabs.harvard.edu/abs/1995MNRAS.274..461B} {274, 461}

\bibitem[\protect\citeauthoryear{{Breivik}, {Chatterjee}  \& {Larson}}{{Breivik} et~al.}{2017}]{Breivik2017}
{Breivik} K.,  {Chatterjee} S.,   {Larson} S.~L.,  2017, \mn@doi [\apjl] {10.3847/2041-8213/aa97d5}, \href {https://ui.adsabs.harvard.edu/abs/2017ApJ...850L..13B} {850, L13}

\bibitem[\protect\citeauthoryear{{Brott} et~al.,}{{Brott} et~al.}{2011}]{Brott2011}
{Brott} I.,  et~al., 2011, \mn@doi [\aap] {10.1051/0004-6361/201016113}, \href {https://ui.adsabs.harvard.edu/abs/2011A&A...530A.115B} {530, A115}

\bibitem[\protect\citeauthoryear{{Chawla}, {Chatterjee}, {Breivik}, {Moorthy}, {Andrews}  \& {Sanderson}}{{Chawla} et~al.}{2022}]{Chawla2022}
{Chawla} C.,  {Chatterjee} S.,  {Breivik} K.,  {Moorthy} C.~K.,  {Andrews} J.~J.,   {Sanderson} R.~E.,  2022, \mn@doi [\apj] {10.3847/1538-4357/ac60a5}, \href {https://ui.adsabs.harvard.edu/abs/2022ApJ...931..107C} {931, 107}

\bibitem[\protect\citeauthoryear{{Choi}, {Dotter}, {Conroy}, {Cantiello}, {Paxton}  \& {Johnson}}{{Choi} et~al.}{2016}]{Choi2016}
{Choi} J.,  {Dotter} A.,  {Conroy} C.,  {Cantiello} M.,  {Paxton} B.,   {Johnson} B.~D.,  2016, \mn@doi [\apj] {10.3847/0004-637X/823/2/102}, \href {https://ui.adsabs.harvard.edu/abs/2016ApJ...823..102C} {823, 102}

\bibitem[\protect\citeauthoryear{{Chun}, {Yoon}, {Jung}, {Kim}  \& {Kim}}{{Chun} et~al.}{2018}]{Chun2018}
{Chun} S.-H.,  {Yoon} S.-C.,  {Jung} M.-K.,  {Kim} D.~U.,   {Kim} J.,  2018, \mn@doi [\apj] {10.3847/1538-4357/aa9a37}, \href {https://ui.adsabs.harvard.edu/abs/2018ApJ...853...79C} {853, 79}

\bibitem[\protect\citeauthoryear{{Cropper} et~al.,}{{Cropper} et~al.}{2018}]{Cropper2018}
{Cropper} M.,  et~al., 2018, \mn@doi [\aap] {10.1051/0004-6361/201832763}, \href {https://ui.adsabs.harvard.edu/abs/2018A&A...616A...5C} {616, A5}

\bibitem[\protect\citeauthoryear{{Di Carlo}, {Agrawal}, {Rodriguez}  \& {Breivik}}{{Di Carlo} et~al.}{2023}]{DiCarlo2023}
{Di Carlo} U.~N.,  {Agrawal} P.,  {Rodriguez} C.~L.,   {Breivik} K.,  2023, \mn@doi [arXiv e-prints] {10.48550/arXiv.2306.13121}, \href {https://ui.adsabs.harvard.edu/abs/2023arXiv230613121D} {p. arXiv:2306.13121}

\bibitem[\protect\citeauthoryear{{Dodd}, {Callingham}, {Helmi}, {Matsuno}, {Ruiz-Lara}, {Balbinot}  \& {L{\"o}vdal}}{{Dodd} et~al.}{2023}]{Dodd2023}
{Dodd} E.,  {Callingham} T.~M.,  {Helmi} A.,  {Matsuno} T.,  {Ruiz-Lara} T.,  {Balbinot} E.,   {L{\"o}vdal} S.,  2023, \mn@doi [\aap] {10.1051/0004-6361/202244546}, \href {https://ui.adsabs.harvard.edu/abs/2023A&A...670L...2D} {670, L2}

\bibitem[\protect\citeauthoryear{{Eggleton}}{{Eggleton}}{1983}]{Eggleton1983}
{Eggleton} P.~P.,  1983, \mn@doi [\apj] {10.1086/160960}, \href {https://ui.adsabs.harvard.edu/abs/1983ApJ...268..368E} {268, 368}

\bibitem[\protect\citeauthoryear{{El-Badry} et~al.,}{{El-Badry} et~al.}{2023a}]{El-Badry2023_bh1}
{El-Badry} K.,  et~al., 2023a, \mn@doi [\mnras] {10.1093/mnras/stac3140}, \href {https://ui.adsabs.harvard.edu/abs/2023MNRAS.518.1057E} {518, 1057}

\bibitem[\protect\citeauthoryear{{El-Badry} et~al.,}{{El-Badry} et~al.}{2023b}]{El-Badry2023_bh2}
{El-Badry} K.,  et~al., 2023b, \mn@doi [\mnras] {10.1093/mnras/stad799}, \href {https://ui.adsabs.harvard.edu/abs/2023MNRAS.521.4323E} {521, 4323}

\bibitem[\protect\citeauthoryear{{El-Badry} et~al.,}{{El-Badry} et~al.}{2024a}]{El-Badry2024_nsall}
{El-Badry} K.,  et~al., 2024a, \mn@doi [arXiv e-prints] {10.48550/arXiv.2405.00089}, \href {https://ui.adsabs.harvard.edu/abs/2024arXiv240500089E} {p. arXiv:2405.00089}

\bibitem[\protect\citeauthoryear{{El-Badry} et~al.,}{{El-Badry} et~al.}{2024b}]{El-Badry2024_ns1}
{El-Badry} K.,  et~al., 2024b, \mn@doi [The Open Journal of Astrophysics] {10.33232/001c.116675}, \href {https://ui.adsabs.harvard.edu/abs/2024OJAp....7E..27E} {7, 27}

\bibitem[\protect\citeauthoryear{{Fern{\'a}ndez}, {Quataert}, {Kashiyama}  \& {Coughlin}}{{Fern{\'a}ndez} et~al.}{2018}]{Fernandez2018}
{Fern{\'a}ndez} R.,  {Quataert} E.,  {Kashiyama} K.,   {Coughlin} E.~R.,  2018, \mn@doi [\mnras] {10.1093/mnras/sty306}, \href {https://ui.adsabs.harvard.edu/abs/2018MNRAS.476.2366F} {476, 2366}

\bibitem[\protect\citeauthoryear{{Gaia Collaboration} et~al.,}{{Gaia Collaboration} et~al.}{2016}]{GaiaCollaboration2016}
{Gaia Collaboration} et~al., 2016, \mn@doi [\aap] {10.1051/0004-6361/201629272}, \href {https://ui.adsabs.harvard.edu/abs/2016A&A...595A...1G} {595, A1}

\bibitem[\protect\citeauthoryear{{Gaia Collaboration} et~al.,}{{Gaia Collaboration} et~al.}{2024}]{Panuzzo2024}
{Gaia Collaboration} et~al., 2024, arXiv e-prints, \href {https://ui.adsabs.harvard.edu/abs/2024arXiv240410486G} {p. arXiv:2404.10486}

\bibitem[\protect\citeauthoryear{{Gieles}, {Erkal}, {Antonini}, {Balbinot}  \& {Pe{\~n}arrubia}}{{Gieles} et~al.}{2021}]{Gieles2021}
{Gieles} M.,  {Erkal} D.,  {Antonini} F.,  {Balbinot} E.,   {Pe{\~n}arrubia} J.,  2021, \mn@doi [Nature Astronomy] {10.1038/s41550-021-01392-2}, \href {https://ui.adsabs.harvard.edu/abs/2021NatAs...5..957G} {5, 957}

\bibitem[\protect\citeauthoryear{{Goldberg} \& {Bildsten}}{{Goldberg} \& {Bildsten}}{2020}]{Goldberg2020}
{Goldberg} J.~A.,  {Bildsten} L.,  2020, \mn@doi [\apjl] {10.3847/2041-8213/ab9300}, \href {https://ui.adsabs.harvard.edu/abs/2020ApJ...895L..45G} {895, L45}

\bibitem[\protect\citeauthoryear{{Hamann}, {Koesterke}  \& {Wessolowski}}{{Hamann} et~al.}{1995}]{Hamann1995}
{Hamann} W.~R.,  {Koesterke} L.,   {Wessolowski} U.,  1995, \aap, \href {https://ui.adsabs.harvard.edu/abs/1995A&A...299..151H} {299, 151}

\bibitem[\protect\citeauthoryear{{Heger}, {Langer}  \& {Woosley}}{{Heger} et~al.}{2000}]{Heger2000}
{Heger} A.,  {Langer} N.,   {Woosley} S.~E.,  2000, \mn@doi [\apj] {10.1086/308158}, \href {https://ui.adsabs.harvard.edu/abs/2000ApJ...528..368H} {528, 368}

\bibitem[\protect\citeauthoryear{{Henyey}, {Vardya}  \& {Bodenheimer}}{{Henyey} et~al.}{1965}]{Henyey1965}
{Henyey} L.,  {Vardya} M.~S.,   {Bodenheimer} P.,  1965, \mn@doi [\apj] {10.1086/148357}, \href {https://ui.adsabs.harvard.edu/abs/1965ApJ...142..841H} {142, 841}

\bibitem[\protect\citeauthoryear{{Holl} et~al.,}{{Holl} et~al.}{2023}]{Holl2023}
{Holl} B.,  et~al., 2023, \mn@doi [\aap] {10.1051/0004-6361/202244161}, \href {https://ui.adsabs.harvard.edu/abs/2023A&A...674A..10H} {674, A10}

\bibitem[\protect\citeauthoryear{{Hurley}, {Tout}  \& {Pols}}{{Hurley} et~al.}{2002}]{Hurley2002}
{Hurley} J.~R.,  {Tout} C.~A.,   {Pols} O.~R.,  2002, \mn@doi [\mnras] {10.1046/j.1365-8711.2002.05038.x}, \href {https://ui.adsabs.harvard.edu/abs/2002MNRAS.329..897H} {329, 897}

\bibitem[\protect\citeauthoryear{{Iorio} et~al.,}{{Iorio} et~al.}{2024}]{Iorio2024}
{Iorio} G.,  et~al., 2024, \mn@doi [arXiv e-prints] {10.48550/arXiv.2404.17568}, \href {https://ui.adsabs.harvard.edu/abs/2024arXiv240417568I} {p. arXiv:2404.17568}

\bibitem[\protect\citeauthoryear{{Jermyn} et~al.,}{{Jermyn} et~al.}{2023}]{Jermyn2023}
{Jermyn} A.~S.,  et~al., 2023, \mn@doi [\apjs] {10.3847/1538-4365/acae8d}, \href {https://ui.adsabs.harvard.edu/abs/2023ApJS..265...15J} {265, 15}

\bibitem[\protect\citeauthoryear{{Klencki}, {Nelemans}, {Istrate}  \& {Pols}}{{Klencki} et~al.}{2020}]{Klencki2020}
{Klencki} J.,  {Nelemans} G.,  {Istrate} A.~G.,   {Pols} O.,  2020, \mn@doi [\aap] {10.1051/0004-6361/202037694}, \href {https://ui.adsabs.harvard.edu/abs/2020A&A...638A..55K} {638, A55}

\bibitem[\protect\citeauthoryear{{Klencki}, {Nelemans}, {Istrate}  \& {Chruslinska}}{{Klencki} et~al.}{2021}]{Klencki2021}
{Klencki} J.,  {Nelemans} G.,  {Istrate} A.~G.,   {Chruslinska} M.,  2021, \mn@doi [\aap] {10.1051/0004-6361/202038707}, \href {https://ui.adsabs.harvard.edu/abs/2021A&A...645A..54K} {645, A54}

\bibitem[\protect\citeauthoryear{{Kotko}, {Banerjee}  \& {Belczynski}}{{Kotko} et~al.}{2024}]{Kotko2024}
{Kotko} I.,  {Banerjee} S.,   {Belczynski} K.,  2024, \mn@doi [arXiv e-prints] {10.48550/arXiv.2403.13579}, \href {https://ui.adsabs.harvard.edu/abs/2024arXiv240313579K} {p. arXiv:2403.13579}

\bibitem[\protect\citeauthoryear{{Kremer}, {Ye}, {Chatterjee}, {Rodriguez}  \& {Rasio}}{{Kremer} et~al.}{2018}]{Kremer2018}
{Kremer} K.,  {Ye} C.~S.,  {Chatterjee} S.,  {Rodriguez} C.~L.,   {Rasio} F.~A.,  2018, \mn@doi [\apjl] {10.3847/2041-8213/aab26c}, \href {https://ui.adsabs.harvard.edu/abs/2018ApJ...855L..15K} {855, L15}

\bibitem[\protect\citeauthoryear{{Langer}, {Fricke}  \& {Sugimoto}}{{Langer} et~al.}{1983}]{Langer1983}
{Langer} N.,  {Fricke} K.~J.,   {Sugimoto} D.,  1983, \aap, \href {https://ui.adsabs.harvard.edu/abs/1983A&A...126..207L} {126, 207}

\bibitem[\protect\citeauthoryear{{Lindegren} et~al.,}{{Lindegren} et~al.}{2018}]{Lindegren2018}
{Lindegren} L.,  et~al., 2018, \mn@doi [\aap] {10.1051/0004-6361/201832727}, \href {https://ui.adsabs.harvard.edu/abs/2018A&A...616A...2L} {616, A2}

\bibitem[\protect\citeauthoryear{{Lindegren} et~al.,}{{Lindegren} et~al.}{2021}]{Lindegren2021}
{Lindegren} L.,  et~al., 2021, \mn@doi [\aap] {10.1051/0004-6361/202039709}, \href {https://ui.adsabs.harvard.edu/abs/2021A&A...649A...2L} {649, A2}

\bibitem[\protect\citeauthoryear{{Maeder}}{{Maeder}}{1987}]{Maeder1987}
{Maeder} A.,  1987, \aap, \href {https://ui.adsabs.harvard.edu/abs/1987A&A...178..159M} {178, 159}

\bibitem[\protect\citeauthoryear{{Majewski} et~al.,}{{Majewski} et~al.}{2017}]{Majewski2017}
{Majewski} S.~R.,  et~al., 2017, \mn@doi [\aj] {10.3847/1538-3881/aa784d}, \href {https://ui.adsabs.harvard.edu/abs/2017AJ....154...94M} {154, 94}

\bibitem[\protect\citeauthoryear{{Mar{\'\i}n Pina}, {Rastello}, {Gieles}, {Kremer}, {Fitzgerald}  \& {Rando}}{{Mar{\'\i}n Pina} et~al.}{2024}]{MarinPina2024}
{Mar{\'\i}n Pina} D.,  {Rastello} S.,  {Gieles} M.,  {Kremer} K.,  {Fitzgerald} L.,   {Rando} B.,  2024, \mn@doi [arXiv e-prints] {10.48550/arXiv.2404.13036}, \href {https://ui.adsabs.harvard.edu/abs/2024arXiv240413036M} {p. arXiv:2404.13036}

\bibitem[\protect\citeauthoryear{{Nieuwenhuijzen} \& {de Jager}}{{Nieuwenhuijzen} \& {de Jager}}{1990}]{Nieuwenhuijzen1990}
{Nieuwenhuijzen} H.,  {de Jager} C.,  1990, \aap, \href {https://ui.adsabs.harvard.edu/abs/1990A&A...231..134N} {231, 134}

\bibitem[\protect\citeauthoryear{{Paxton}, {Bildsten}, {Dotter}, {Herwig}, {Lesaffre}  \& {Timmes}}{{Paxton} et~al.}{2011}]{Paxton2011}
{Paxton} B.,  {Bildsten} L.,  {Dotter} A.,  {Herwig} F.,  {Lesaffre} P.,   {Timmes} F.,  2011, \mn@doi [\apjs] {10.1088/0067-0049/192/1/3}, \href {https://ui.adsabs.harvard.edu/abs/2011ApJS..192....3P} {192, 3}

\bibitem[\protect\citeauthoryear{{Paxton} et~al.,}{{Paxton} et~al.}{2013}]{Paxton2013}
{Paxton} B.,  et~al., 2013, \mn@doi [\apjs] {10.1088/0067-0049/208/1/4}, \href {https://ui.adsabs.harvard.edu/abs/2013ApJS..208....4P} {208, 4}

\bibitem[\protect\citeauthoryear{{Paxton} et~al.,}{{Paxton} et~al.}{2015}]{Paxton2015}
{Paxton} B.,  et~al., 2015, \mn@doi [\apjs] {10.1088/0067-0049/220/1/15}, \href {https://ui.adsabs.harvard.edu/abs/2015ApJS..220...15P} {220, 15}

\bibitem[\protect\citeauthoryear{{Paxton} et~al.,}{{Paxton} et~al.}{2018}]{Paxton2018}
{Paxton} B.,  et~al., 2018, \mn@doi [\apjs] {10.3847/1538-4365/aaa5a8}, \href {https://ui.adsabs.harvard.edu/abs/2018ApJS..234...34P} {234, 34}

\bibitem[\protect\citeauthoryear{{Paxton} et~al.,}{{Paxton} et~al.}{2019}]{Paxton2019}
{Paxton} B.,  et~al., 2019, \mn@doi [\apjs] {10.3847/1538-4365/ab2241}, \href {https://ui.adsabs.harvard.edu/abs/2019ApJS..243...10P} {243, 10}

\bibitem[\protect\citeauthoryear{{Rastello}, {Iorio}, {Mapelli}, {Arca-Sedda}, {Di Carlo}, {Escobar}, {Shenar}  \& {Torniamenti}}{{Rastello} et~al.}{2023}]{Rastello2023}
{Rastello} S.,  {Iorio} G.,  {Mapelli} M.,  {Arca-Sedda} M.,  {Di Carlo} U.~N.,  {Escobar} G.~J.,  {Shenar} T.,   {Torniamenti} S.,  2023, \mn@doi [\mnras] {10.1093/mnras/stad2757}, \href {https://ui.adsabs.harvard.edu/abs/2023MNRAS.526..740R} {526, 740}

\bibitem[\protect\citeauthoryear{{Romagnolo}, {Belczynski}, {Klencki}, {Agrawal}, {Shenar}  \& {Sz{\'e}csi}}{{Romagnolo} et~al.}{2023}]{Romagnolo2023}
{Romagnolo} A.,  {Belczynski} K.,  {Klencki} J.,  {Agrawal} P.,  {Shenar} T.,   {Sz{\'e}csi} D.,  2023, \mn@doi [\mnras] {10.1093/mnras/stad2366}, \href {https://ui.adsabs.harvard.edu/abs/2023MNRAS.525..706R} {525, 706}

\bibitem[\protect\citeauthoryear{{Schootemeijer}, {Langer}, {Grin}  \& {Wang}}{{Schootemeijer} et~al.}{2019}]{Schootemeijer2019}
{Schootemeijer} A.,  {Langer} N.,  {Grin} N.~J.,   {Wang} C.,  2019, \mn@doi [\aap] {10.1051/0004-6361/201935046}, \href {https://ui.adsabs.harvard.edu/abs/2019A&A...625A.132S} {625, A132}

\bibitem[\protect\citeauthoryear{{Tanikawa}, {Wang}  \& {Fujii}}{{Tanikawa} et~al.}{2024a}]{Tanikawa2024b}
{Tanikawa} A.,  {Wang} L.,   {Fujii} M.~S.,  2024a, \mn@doi [arXiv e-prints] {10.48550/arXiv.2404.01731}, \href {https://ui.adsabs.harvard.edu/abs/2024arXiv240401731T} {p. arXiv:2404.01731}

\bibitem[\protect\citeauthoryear{{Tanikawa}, {Cary}, {Shikauchi}, {Wang}  \& {Fujii}}{{Tanikawa} et~al.}{2024b}]{Tanikawa2024}
{Tanikawa} A.,  {Cary} S.,  {Shikauchi} M.,  {Wang} L.,   {Fujii} M.~S.,  2024b, \mn@doi [\mnras] {10.1093/mnras/stad3294}, \href {https://ui.adsabs.harvard.edu/abs/2024MNRAS.527.4031T} {527, 4031}

\bibitem[\protect\citeauthoryear{{Tauris} \& {van den Heuvel}}{{Tauris} \& {van den Heuvel}}{2023}]{Tauris2023}
{Tauris} T.~M.,  {van den Heuvel} E. P.~J.,  2023, {Physics of Binary Star Evolution. From Stars to X-ray Binaries and Gravitational Wave Sources}, \mn@doi{10.48550/arXiv.2305.09388.
}

\bibitem[\protect\citeauthoryear{{Tauris} et~al.,}{{Tauris} et~al.}{2017}]{Tauris2017}
{Tauris} T.~M.,  et~al., 2017, \mn@doi [\apj] {10.3847/1538-4357/aa7e89}, \href {https://ui.adsabs.harvard.edu/abs/2017ApJ...846..170T} {846, 170}

\bibitem[\protect\citeauthoryear{{Thielemann}, {Nomoto}  \& {Hashimoto}}{{Thielemann} et~al.}{1996}]{Thielemann1996}
{Thielemann} F.-K.,  {Nomoto} K.,   {Hashimoto} M.-A.,  1996, \mn@doi [\apj] {10.1086/176980}, \href {https://ui.adsabs.harvard.edu/abs/1996ApJ...460..408T} {460, 408}

\bibitem[\protect\citeauthoryear{{Vink}, {de Koter}  \& {Lamers}}{{Vink} et~al.}{2000}]{Vink2000}
{Vink} J.~S.,  {de Koter} A.,   {Lamers} H.~J.~G.~L.~M.,  2000, \mn@doi [\aap] {10.48550/arXiv.astro-ph/0008183}, \href {https://ui.adsabs.harvard.edu/abs/2000A&A...362..295V} {362, 295}

\bibitem[\protect\citeauthoryear{{Vink}, {de Koter}  \& {Lamers}}{{Vink} et~al.}{2001}]{Vink2001}
{Vink} J.~S.,  {de Koter} A.,   {Lamers} H.~J.~G.~L.~M.,  2001, \mn@doi [\aap] {10.1051/0004-6361:20010127}, \href {https://ui.adsabs.harvard.edu/abs/2001A&A...369..574V} {369, 574}

\bibitem[\protect\citeauthoryear{{Woosley} \& {Heger}}{{Woosley} \& {Heger}}{2006}]{Woosley2006}
{Woosley} S.~E.,  {Heger} A.,  2006, \mn@doi [\apj] {10.1086/498500}, \href {https://ui.adsabs.harvard.edu/abs/2006ApJ...637..914W} {637, 914}

\bibitem[\protect\citeauthoryear{{Woosley}, {Heger}  \& {Weaver}}{{Woosley} et~al.}{2002}]{Woosley2002}
{Woosley} S.~E.,  {Heger} A.,   {Weaver} T.~A.,  2002, \mn@doi [Reviews of Modern Physics] {10.1103/RevModPhys.74.1015}, \href {https://ui.adsabs.harvard.edu/abs/2002RvMP...74.1015W} {74, 1015}

\bibitem[\protect\citeauthoryear{{Yoon} \& {Langer}}{{Yoon} \& {Langer}}{2005}]{Yoon2005}
{Yoon} S.~C.,  {Langer} N.,  2005, \mn@doi [\aap] {10.1051/0004-6361:20054030}, \href {https://ui.adsabs.harvard.edu/abs/2005A&A...443..643Y} {443, 643}

\bibitem[\protect\citeauthoryear{{Yoon}, {Langer}  \& {Norman}}{{Yoon} et~al.}{2006}]{Yoon2006}
{Yoon} S.~C.,  {Langer} N.,   {Norman} C.,  2006, \mn@doi [\aap] {10.1051/0004-6361:20065912}, \href {https://ui.adsabs.harvard.edu/abs/2006A&A...460..199Y} {460, 199}

\makeatother
\end{thebibliography}



\end{document}